\documentclass[prl,nofootinbib,superscriptaddress,twocolumn]{revtex4-1}
\usepackage[a4paper,left=1.5cm,right=1.5cm,top=3cm,bottom=3cm]{geometry}

\usepackage{verbatim}
\usepackage{color}
\usepackage{amsfonts,amssymb,mathrsfs,amsmath}
\usepackage{framed}
\usepackage{mdframed}
\usepackage{latexsym}
\usepackage{graphicx}
\usepackage{datetime}
\newdateformat{mydate}{\THEDAY{ }\monthname[\THEMONTH]{ }\THEYEAR}

\allowdisplaybreaks

\usepackage{tikz}
\usepackage{color}
\usepackage{framed}
\usepackage{hyperref}
\hypersetup{colorlinks, citecolor=bluscuro, linkcolor=black, urlcolor=bluscuro}
\definecolor{rossos}{cmyk}{0,1,1,0.55}
\definecolor{bluscuro}{rgb}{0.15, 0.2, .85}
\definecolor{bluchiaro}{cmyk}{1,.3,0.,0.1}

\newcommand{\be}{\begin{equation}}
\newcommand{\ee}{\end{equation}}
\renewcommand{\d}{{\rm d}}

\def\PBH{\text{\tiny PBH}}

\def\g{\text{\tiny g}}
\def\NL{\text{\tiny NL}}

\def\vx{{\vec{x}}}

\def\lsim{\mathrel{\rlap{\lower4pt\hbox{\hskip0.5pt$\sim$}}
    \raise1pt\hbox{$<$}}}         
\def\gsim{\mathrel{\rlap{\lower4pt\hbox{\hskip0.5pt$\sim$}}
    \raise1pt\hbox{$>$}}}         

\newcommand{\arXiv}[2]{\href{http://arxiv.org/pdf/#1}{{\tt [#2/#1]}}}
\newcommand{\arXivold}[1]{\href{http://arxiv.org/pdf/#1}{{\tt [#1]}}}

\begin{document}

\title{Constraining the Initial Primordial Black Hole Clustering with  CMB-distortion}

\author{V. De Luca}
\email{Valerio.DeLuca@unige.ch}
\address{D\'epartement de Physique Th\'eorique and Centre for Astroparticle Physics (CAP), Universit\'e de Gen\`eve, 24 quai E. Ansermet, CH-1211 Geneva, Switzerland}
\address{Dipartimento di Fisica, “Sapienza” Universit\`a di Roma, Piazzale Aldo Moro 5, 00185, Roma, Italy}

\author{G. Franciolini}
\email{Gabriele.Franciolini@unige.ch}
\address{D\'epartement de Physique Th\'eorique and Centre for Astroparticle Physics (CAP), Universit\'e de Gen\`eve, 24 quai E. Ansermet, CH-1211 Geneva, Switzerland}

\author{A.~Riotto}
\email{Antonio.Riotto@unige.ch}
\address{D\'epartement de Physique Th\'eorique and Centre for Astroparticle Physics (CAP), Universit\'e de Gen\`eve, 24 quai E. Ansermet, CH-1211 Geneva, Switzerland}

\date{\today}

\begin{abstract}
\noindent
 The merger rate of primordial black holes depends on their initial clustering. In the absence of primordial non-Gaussianity  correlating short and large-scales,  primordial black holes are distributed \`a la Poisson at the time of their formation. However,  primordial non-Gaussianity of the local-type may correlate primordial black holes on large-scales. We show that future experiments looking for  CMB $\mu$-distortion  would test the hypothesis of  initial primordial black hole clustering induced by local non-Gaussianity,  while existing limits already show that significant non-Gaussianity is necessary to induce primordial black hole clustering.
\end{abstract}

\maketitle

\vskip 0.5cm
\noindent
\paragraph{Introduction.}
\noindent
The  LIGO/Virgo Collaboration has by now reported 
several detections of  Gravitational Waves~(GWs) coming from black hole~(BH) mergers ~\cite{LIGOScientific:2018mvr, Abbott:2020niy}.  
Several studies have developed the description of
Primordial Black Holes (PBHs) binary formation and merger rates~\cite{Nakamura:1997sm,Ioka:1998nz,Sasaki:2016jop,Bird:2016dcv,Clesse:2016vqa, Wang:2016ana, Ali-Haimoud:2017rtz, 
Chen:2018czv, Kavanagh:2018ggo,raidal,ver, Gow:2019pok,  Fernandez:2019kyb,Hall:2020daa,DeLuca:2020sae, Wong:2020yig, DeLuca:2020jug,Hutsi:2020sol}.
Interestingly, current data allow for a fraction of the observed events to be PBHs~\cite{us,Franciolini:2021tla}.

In the absence of primordial non-Gaussianity (NG), PBHs are initially predominantly Poisson distributed
 (meaning that the most sizeable contribution to the PBH correlation function at the relevant scales comes from the Poisson noise typical of discrete tracers)~\cite{Ali-Haimoud:2018dau, in, Ballesteros:2018swv, MoradinezhadDizgah:2019wjf}
and the corresponding merger rate   allows the fraction $f_\PBH$ of PBHs  to the dark matter to be below
the percent level~\cite{Wong:2020yig}.   Clustering  at the time of formation of  PBHs can crucially affect the present and past merger rate of PBH binaries, both by boosting the formation of binaries and enhancing the subsequent  potential suppression due to interaction of binaries in PBH clusters. In particular, the latter effect was advocated in the literature to possibly allow for larger values of $f_\PBH$ and therefore a major role of PBHs in the dark matter budget~\cite{Raidal:2017mfl, Young:2019gfc, atal}. 

\begin{figure}[t!]
	\centering
	\includegraphics[width=0.49\textwidth]{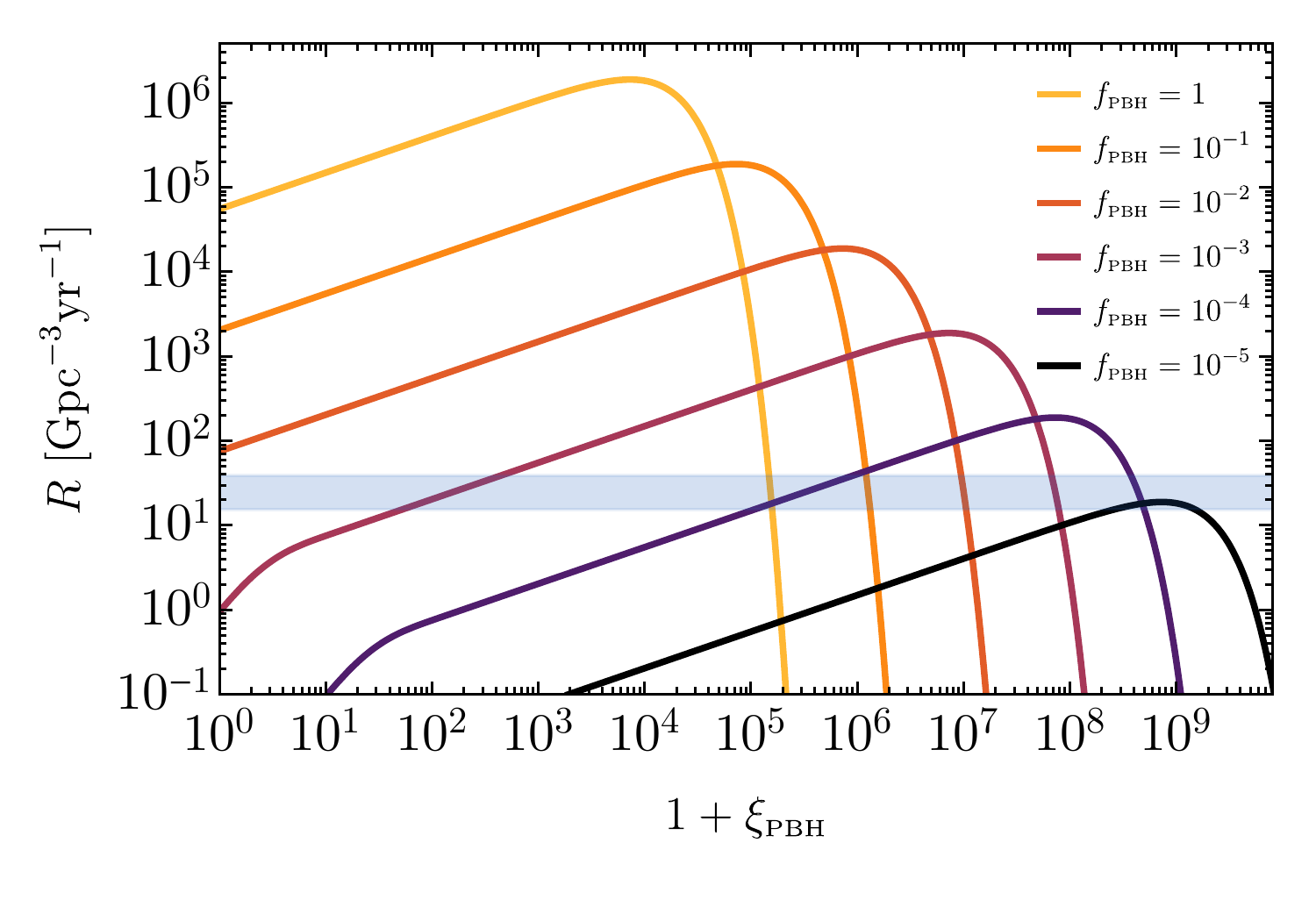}
	\caption{\it The upper bound of the PBH binary merger rate today as a function of a constant PBH correlation $\xi_\PBH$, for different values of the PBH abundance and for a PBH mass of 30 $M_\odot$. The shaded region indicates the LIGO/Virgo current detection band, which cannot be reached for $f_\text{\tiny \rm PBH}\lesssim 10^{-5}$ even  when PBHs are clustered at formation.
	}
	\label{figrate}
\end{figure}

Primordial NG  (of the local type) allows for a cross-talk between small and large-scales~\cite{ng},  correlating the horizon-size regions where the PBHs are initially formed upon collapse of  the large overdensities  generated during inflation, see Refs.~\cite{sasaki, Green:2020jor} for recent reviews. PBHs may be  therefore clustered in the presence of local NG. 
If clustering and $f_\PBH$ are large enough, then the initial typical distance between two PBHs becomes so small that mergers occur at epochs earlier than the 
current age of the universe, making the corresponding GWs not detectable by  the LIGO/Virgo Collaboration. This is reflected 
by the fact that the upper bound (that is, not accounting for the  dynamical suppression due to the binary disruption in small structures~\cite{ver,Jedamzik:2020ypm,DeLuca:2020jug}) of the merger rate today $R_4\equiv R/(10^4{\rm Gpc}^{-3} {\rm yr}^{-1})$  as a function of the  PBH correlation $\delta_\text{\tiny dc} = 1+\xi_\PBH$ (up to the binary scales) goes like~\cite{Raidal:2017mfl} 
\begin{equation}
	 R_4\simeq 
	\begin{cases}
	1.5 \cdot 10^5 \,
	\delta_\text{\tiny dc}^2
	f_\PBH^{3},
	\hfill  \,\, \text{for} \,\, 
	\delta_\text{\tiny dc} f_\PBH \,\,\lesssim 7\cdot 10^{-3},
	\\[7pt] 
	5.5\,\delta_\text{\tiny dc}^{16/37} 
	f_\PBH^{53/37}, 
	\hfill  \,\, \text{for} \,\, 
	7\cdot 10^{-3} \lsim 	\delta_\text{\tiny dc} f_\PBH \lsim 10^3,
	\\[7pt] 
		 0.8\,
		 \delta_\text{\tiny dc}^{0.7} f_\PBH^{1.7}
		 e^{-  \delta_\text{\tiny dc} f_\PBH /10^4 },
		 			\hfill  \,\, \text{for} \,\, 
	\delta_\text{\tiny dc} f_\PBH \gsim 10^3,
	\end{cases}
\end{equation}
and, therefore, the merger rate is exponentially suppressed for $\xi_\PBH f_\PBH \gsim 10^4$, see Fig.~\ref{figrate}. 
 This would already be sufficient to evade the constraints proposed in Ref.~\cite{Bringmann:2018mxj} on clustered PBH scenarios, which are, however, not accounting for the dynamical suppression of the merger rate.
{ Fig.~\ref{figrate} is useful to understand the generic impact of PBH clustering on the merger rate and, as such, we have allowed large values of $\xi_\PBH$, as predicted, for instance, in \cite{atal}. However, as we will see in the following, 
we will be interested in constraining smaller values of the combination between the PBH abundance and the correlation function.}

One should not claim victory too soon, though.   First of all, local NG is currently limited by Planck observations \cite{Planck}. Secondly,  it   introduces a mode-coupling  to   the observed CMB  scales and   a significant dark matter isocurvature mode is introduced as  the number density of PBHs varies  in different regions of the universe on large scales.  For large values of $f_\PBH$ such an isocurvature component  is excluded by the recent Planck data  \cite{tada,by,an}. We will come back to this point later on.

The goal of this paper is to stress that there is another argument one should consider when dealing with a large PBH clustering induced by NG. 
  For the interesting case of PBHs with masses around $\sim   30\, M_\odot$, the range of initial comoving distances relevant for the calculation of the present merger rate is $(4 \cdot 10^{-5}\div 10^{-3})$  Mpc  \cite{sasaki}. Indeed, only PBHs separated by a distance smaller than $\sim 10^{-3}$ Mpc  can form a binary
system, while there is
also a  minimum separation $\sim 4 \cdot 10^{-5}$ Mpc  above  which  PBH binaries undergo mergers within a timescale allowing for the GW signal emitted to be observable at LIGO/Virgo detectors.
 
 This range of scales  strongly overlaps with the interval  where CMB $\mu$-distortion may take place, that is in the range $(10^{-4}\div 2\cdot 10^{-2})$ Mpc,   not accessible from CMB anisotropies observations  (notice that $y$-distortions involve larger comoving scales~\cite{Chluba:2012we} and are therefore not relevant for the scales involved in our arguments).  

We will show that the possibility of enhancing  PBH clustering  through primordial NG, so that the PBH merger rate is significantly altered, may be tested by future measurements of the CMB $\mu$-distortion. 
 
 CMB distortion  is caused by the energy injection originated by the dissipation of acoustic waves through the Silk damping as they re-enter the horizon and start oscillating \cite{s, Chluba:2015bqa}.  
Furthermore,  as PBH clustering is induced by  a sizeable curvature  power spectrum  on the scales relevant for the merger rate, and those scales overlap with those where the CMB is most sensitive to a  large curvature perturbation through $\mu$-distortion, the connection is evident.
  Forecasted constraints from PIXIE ($\mu<3\cdot 10^{-8}$) \cite{mu1},
 from SuperPIXIE ($\mu<7\cdot 10^{-9}$) \cite{mu2}, Voyage2050  ($\mu<1.9\cdot 10^{-9}$) and 10 $\times$ Voyage2050  ($\mu<1.9\cdot 10^{-10}$) \cite{mu3}  would, in case these future experiments will be realised, allow to test the hypothesis of large PBH clustering induced by primordial NG.

\vskip 0.5cm
\noindent
\paragraph{PBH clustering in the presence of primordial NG.} PBHs may form  if the energy density perturbation generated during inflation is sizeable enough.  When after inflation the corresponding wavelengths are re-entering the horizon, 
the large density contrast collapses to form PBHs almost immediately after horizon re-entry \cite{sasaki, 1967SvA....10..602Z, Hawking:1974rv, Chapline:1975ojl, s1, s00}, and the resulting  PBH mass is of the order of the mass  contained in the corresponding horizon volume. Since PBHs are discrete tracers, 
the overdensity of PBHs  reads 
\begin{equation}\label{deltaPBH}
\delta_\PBH({\vec x})=\frac{1}{n_\PBH}\sum_i \delta_D(\vec x-\vec x_i)-1,
\end{equation}
where $\delta_D(\vec x)$ is the three-dimensional Dirac distribution, $n_\PBH \simeq f_\PBH (30 M_\odot/M_\PBH) \, {\rm kpc}^{-3}$ is the average number density of PBHs per comoving volume and $i$ runs over the initial positions of PBHs. The corresponding  two-point correlation function is \cite{in}
\begin{align}\label{deltadeltaPBH}
\big\langle\delta_\PBH(\vec x)\delta_\PBH( 0) \big\rangle 
&= \frac{1}{n_\PBH}\delta_D(\vx)+ \xi_\PBH(x),
\end{align}
in terms of the  Poisson piece and the reduced PBH correlation function $\xi_\PBH(x)$. Notice that $\xi_\PBH(x) \sim 1$ is the benchmark value to have PBHs spatially correlated at initial distances relevant for the calculation of the present merger rate. To characterise the latter and to introduce a sizeable PBH clustering on large-scales, we start from the curvature perturbation $\zeta(\vec x)$ and adopt  the following  generic NG parametrisation \cite{Suyama:2019cst,atal}
\begin{equation}
\zeta({\vec x})=(1+\alpha \chi ({\vec x})) \zeta_{\g} ({\vec x}), \label{p-tnl}
\end{equation}
where $\zeta_{\g}(\vec x)$ is the Gaussian part  of the curvature perturbation.
There are two options at this point, either the  $\chi(\vec x)$ coincides with the  curvature  field itself, $\zeta_\g(\vec x)$, or it does not. 

In the first case, we recover the familiar local-type NG model and $\alpha$ is  the standard $f_\NL$ parameter. We assume that the Gaussian  curvature perturbation has three components, one at short-scales $\sim k_s^{-1}$ responsible for the generation of the PBHs, one at long scales $\sim k_l^{-1}$ at which the PBH clustering is sourced and the standard almost scale-invariant contribution responsible for the CMB anisotropies
\begin{equation}\label{pszeta}
	{\cal P}_{\g}(k) =k_s A_s \delta_\text{\tiny D}(k-k_s)+k_l A_l \delta_\text{\tiny D}(k-k_l) +{\cal P}_{\textrm{\tiny CMB}}(k),
\end{equation}
where we have assumed  a Dirac delta shape for the power spectrum of the curvature perturbation on small (large)-scales with amplitude $A_s (A_l)$. 
In such a case the PBH power spectrum on large-scales $\sim k_l^{-1}$ reads \cite{tada}
\be
\mathcal{P}_{\delta_\PBH} (k) \simeq 
4 \nu^4 f_\NL^2 A_l k_l\delta (k - k_l),
\ee 
where $\nu=(\delta_c/\sigma)$ is the bias factor due to the fact that PBHs are born from peaks of the underlying radiation energy density perturbation and  $\delta_c\simeq 0.59$ is the threshold for PBH formation, see Refs.~\cite{Musco:2018rwt, Germani:2018jgr, Musco}. 
The variance $\sigma^2$ of the density field is given by
\be
\sigma^2 = \frac{16}{81}\int_0^\infty {\rm d}\ln k \,T^2(k, r_m) W^2(k,r_m) (k r_m)^4{\cal P}_	\text{\tiny g}(k),
\ee
as a function of the real space top hat window function $W$, the transfer function $T$ in a radiation dominated universe  and the PBH relevant scale for collapse $r_m = 2.74 / k_s$ \cite{Musco}. 
For a PBH population with mass $M_\PBH \simeq 30 M_\odot$ and abundance $f_\PBH \simeq 10^{-3}$ related to the LIGO/Virgo observations, the relevant short scale spectrum parameters are $k_s \simeq  2.4 \cdot 10^{5} {\rm Mpc}^{-1} $ and $A_s \simeq 0.0063$. Notice that this parameter space is not yet constrained by pulsar timing array experiments \cite{Gow:2020bzo}.
The corresponding initial PBH correlation function is~\cite{Bardeen:1985tr}
\begin{align}\label{NGcorrfunction}
\xi_\PBH(x) = \int_0^\infty \frac{\d k}{k} \mathcal{P}_{\delta_\PBH} (k) \, j_0(kx) 	
\simeq  4 \nu^4 f_\NL^2 A_l j_0 (k_l x),
\end{align}
where $j_0$ identifies the zeroth spherical Bessel function.
In the alternative case in which $\chi(\vec x)$ is not the curvature perturbation $\zeta_\g(\vec x)$, we assume for simplicity that it is not correlated with it and that it possesses  a power spectrum
\be
\label{pchi}
\mathcal{P}_{\chi} (k)=k_l A_l \delta_\text{\tiny D}(k-k_l),
\ee
while the power spectrum of $\zeta_\g(\vec x)$ has only the short-scale piece responsible for PBH formation and the CMB contribution
\begin{equation}\label{pszetachi}
	{\cal P}_{\g}(k) =k_s A_s \delta_\text{\tiny D}(k-k_s)+{\cal P}_{\textrm{\tiny CMB}}(k).
\end{equation}
The resulting initial PBH correlation function  is \cite{Suyama:2019cst}
\begin{equation}
\xi_{\PBH} (x) \approx 	\frac{225}{64} \nu^4 \alpha^2 A_l j_0 (k_l x).
\end{equation}
Notice that the expressions we have presented are valid for $\xi_\PBH\lsim 1$, which will be  consistent with the results found in the coming sections.
Since no  physical process can affect the relative separation $x$ between two PBHs so long as $x$ is larger than the horizon scale, the PBH correlation function does not change
when $k^{-1}_l$ is outside the horizon. Upon horizon re-entry,   the PBH density contrast is essentially frozen until matter-radiation equivalence, and subsequently grows linearly according to \cite{Inman:2019wvr,DeLuca:2020jug}
\be
\label{linear}
\xi_{\PBH} (x,z)\simeq\left(1+ \frac{3}{2}f_\PBH\frac{1+z_\text{{\tiny eq}}}{1+z}\right)^2 \xi_{\PBH} (x),
\ee
in which we adopt the matter-dominated epoch behaviour  $(1+z)^{-1}$ for simplicity and where $z_\text{\tiny eq}$ indicates the 
 redshift at matter-radiation equality.

\begin{figure*}[th!]
\centering
\includegraphics[width=0.49\textwidth]{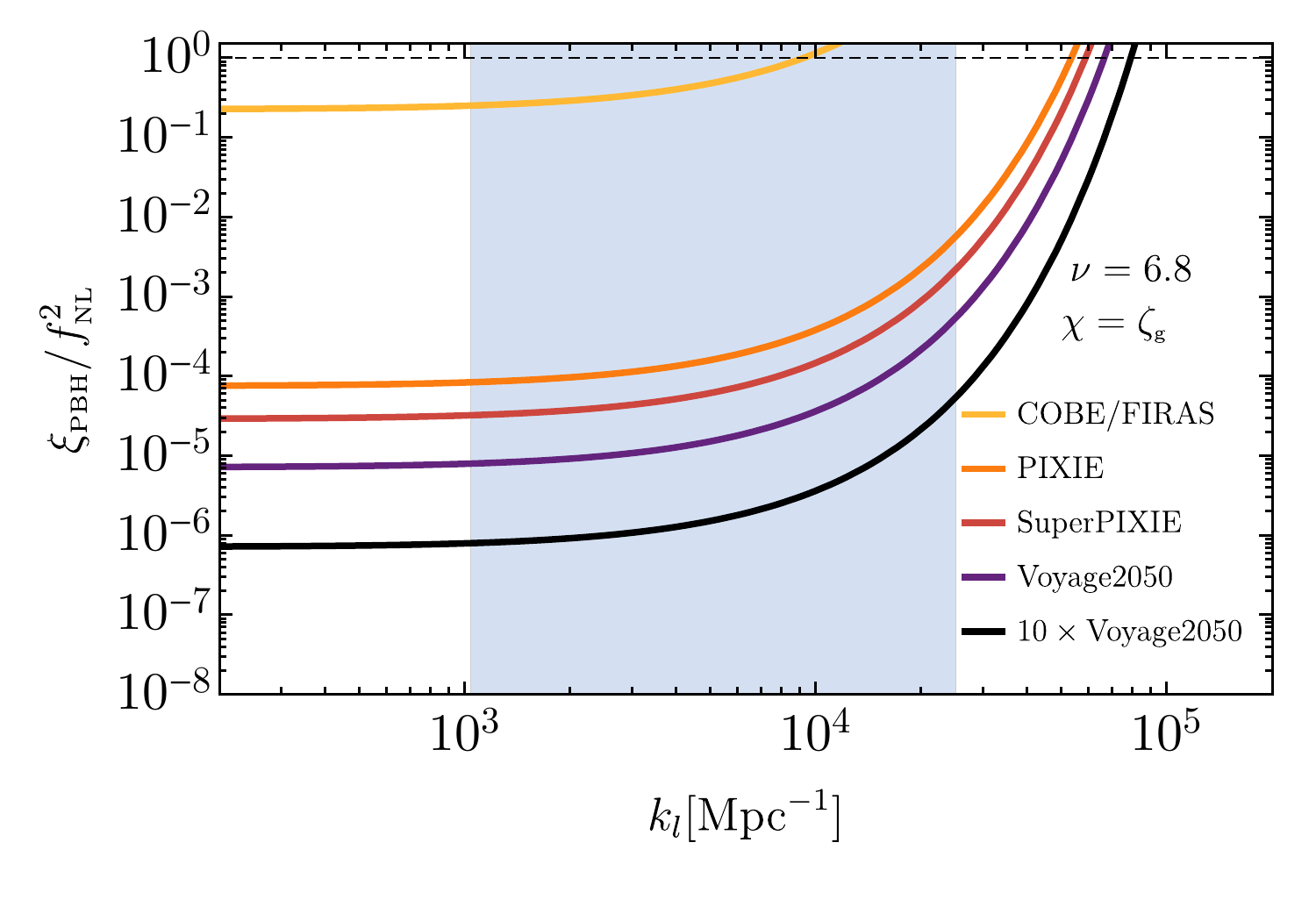}
\includegraphics[width=0.49\textwidth]{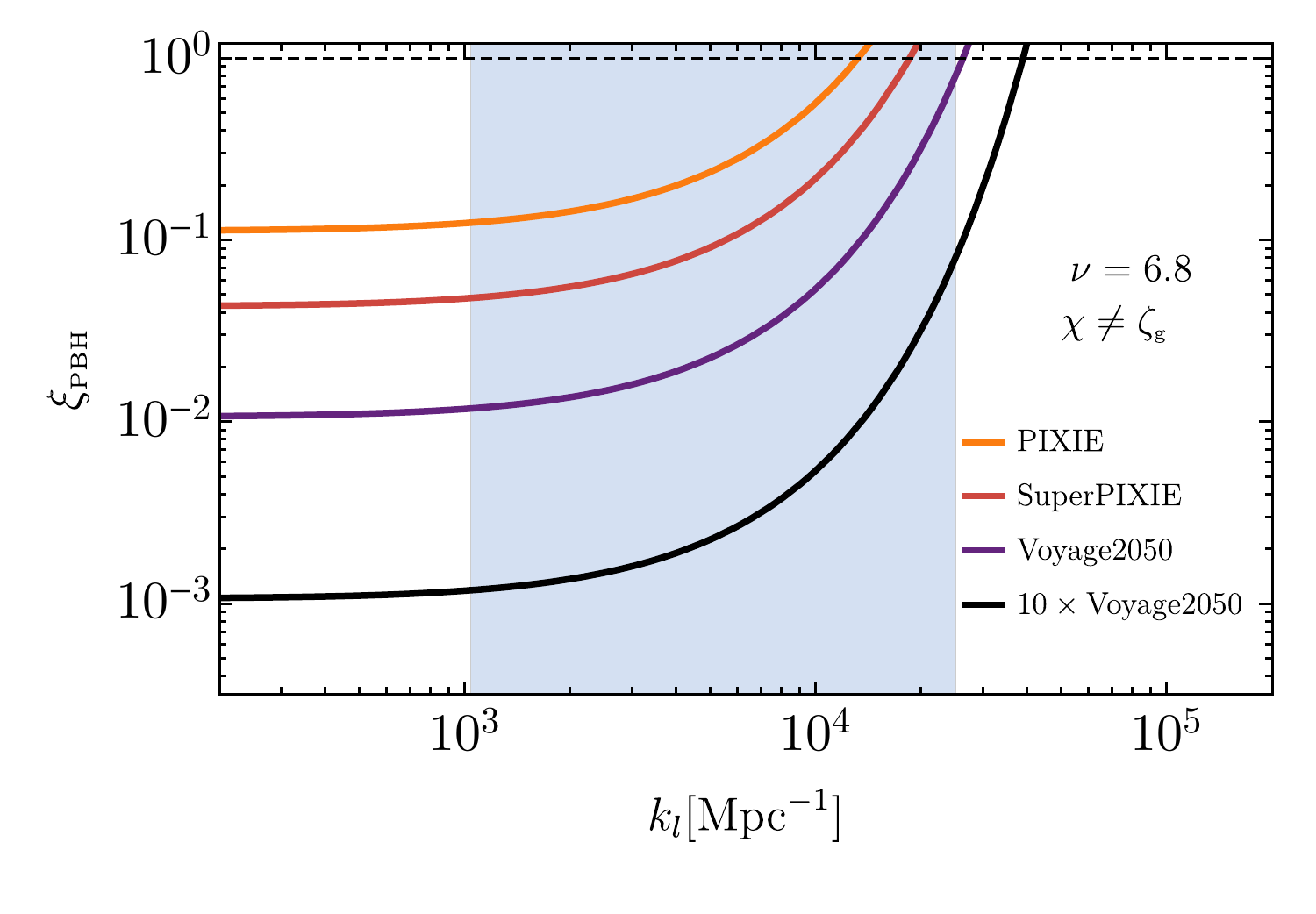}
\caption{\it Limits on the PBH correlation function from the CMB $\mu$-distortion. To fix the value of $\nu\simeq6.8$ we  have chosen the representative value $M_\text{\tiny \rm PBH} = 30 M_\odot$  for the PBH mass,  for which $f_\text{\tiny \rm PBH}=10^{-3}$ in agreement with the current constraints \cite{carr}. The blue band indicates the range of scales relevant for the binary formation.
 }
\label{figcorr}
\end{figure*}

Since the characteristic time for PBH binary formation is before matter-radiation equality, around redshifts $z\sim 10^4$, the correlation function is not expected to change significantly between PBH formation epoch and the binary formation epoch. On the other hand, the corresponding radiation correlation function, the peaks of  which may end up in PBHs, grows as $(1+z)^{-4}$ till the mode $k_l^{-1}$ enters the horizon and afterwards it remains roughly constant in  time. A too large radiation correlation function will correspond to a large energy injection in the system and to a large $\mu$-distortion.
\vskip 0.5cm
\noindent
\paragraph{CMB $\mu$-distortion.}
Silk damping causes the dissipation of acoustic waves  in the photon-baryon plasma, thus  injecting energy into the CMB and causing the CMB spectral distortions.
Following  Refs.~\cite{Hu:1994bz, Dent:2012ne}, the $\mu$-distortion is
\begin{align}\label{mu}
\mu = 1.4\int_{z_1}^{z_2}\d z\frac{\d Q/\d z}{\overline{\rho}_{r}}e^{-(z/z_\text{\tiny DC})^{5/2}}, 
\end{align}
where $z_\text{\tiny DC} 
\simeq 2.6 \cdot 10^6$ is the redshift  scale for double Compton scattering. 
The energy release per unit redshift is given by
\begin{equation}\label{dQ/dz}
\frac{\d Q/\d z}{\overline{\rho}_{r}}=-\int\frac{\d k}{k}{\cal P}_{r}(k,z)\frac{\d \Delta_{Q}^2}{\d z},
\end{equation}
with
 \be
 \Delta_Q^2 (k) = {9 c_s^2 \over 2 } e^{-2 k^2/k_\text{\tiny D}^2},
\ee
in terms of the sound speed $c_s$ and the diffusion scale
\begin{eqnarray}
k_\text{\tiny D} = A_\text{\tiny D}^{-1/2}(1+z)^{3/2},\,\,\,
A_\text{\tiny D} \simeq  6\cdot 10^{10} \,\textrm{Mpc}^{2}.
\end{eqnarray}
The radiation power spectrum is related to the curvature perturbation power spectrum by the standard relation ${\cal P}_{r}(k,a)\simeq
(4/9)^2(k/aH)^4 T^2(k,a){\cal P}_{\zeta}(k)$, where $a$ is the scale factor and $H$ the Hubble rate.  For the relevant large scales,  in the scenario in which $\chi$ coincides with $\zeta_g$, the adopted curvature perturbation power spectrum  directly corresponds to the peaked piece proportional to the large-scale amplitude $A_l$ in Eq.~\eqref{pszeta}. In the alternative scenario when $\chi$ and  $\zeta_g$ are different, the characteristic curvature power spectrum would be given by ${\cal P}_{\zeta}(k) \simeq 25 A_l \alpha^2 A_s^2 \delta_\text{\tiny D} (k-k_l)$. The higher power in the short-scale amplitude $A_s$ comes from the higher order correlations of Eq.~\eqref{p-tnl} needed to connect two distant points and the numerical factor $25$ arises from the corresponding combinatorial counting.

We evaluate the  $\mu$-distortion for the injection interval determined by the double Compton scattering decoupling $z_1 =2\cdot 10^6$ and the thermalization decoupling by Compton scattering $z_2 = 5\cdot 10^4$. Indeed, at $z\gsim z_1$ the content of the universe can be described by a photon-baryon fluid in thermal equilibrium which has a black-body spectrum. This equilibrium is achieved mainly through elastic and double Compton scattering. However, at later times $z\lsim z_2$ double Compton scattering is no longer efficient whereas the single Compton scattering still provides equilibrium.

In the case in which the large-scale field $\chi(\vec x)$ coincides with the curvature perturbation, the $\mu$-distortion is found to be
\be
	\mu  \simeq
	\frac{16}{81} A_l  {\cal I}(k_l) \simeq
	\frac{4}{81} \frac{\xi_{\PBH}}{\nu^4 f_\NL^2}  
	 {\cal I}(k_l), 
	 \ee
	 where
	 \be
	{\cal I}(k_l)= \frac{189}{5} A_\text{\tiny D} k_l^2 c_s^2 \int_{z_2}^{z_1}\frac{\d z}
	{(1+z)^4}
	e^{-(z/z_\text{\tiny DC})^{5/2}}
	e^{-2 k_l ^2/k_\text{\tiny D}^2}.
\ee
In the opposite case, where the $\chi(\vec x)$ does not coincide with the curvature perturbation, we find
\begin{align}
	 \mu   \simeq
	 \frac{2025}{16} A_l \alpha^2  \frac{\delta_c^4}{\nu^4} 
	 {\cal I}(k_l) \simeq 36 \frac{\xi_\PBH \delta_c^4}{\nu^8} 
	 {\cal I}(k_l).
 \end{align}
In both cases  we have assumed the PBH clustering correlation function to be constant for $x\lsim k_l^{-1}$. 
 \vskip 0.5cm
\noindent
\paragraph{Results and conclusions.}
In Fig.~\ref{figcorr} we plot the forecasted limits on the PBH correlation function at the scales relevant for the merger rate coming from the CMB $\mu$-distortion. 

In the standard
$f_\NL$ local-type NG, the distortion is directly proportional to the amplitude $A_l$ of the large-scale part of the curvature perturbation and therefore only a large value of $f_\NL$
may provide a PBH correlation $\xi_\PBH\gsim 1$. For instance, if PIXIE does not find any CMB $\mu$-distortion,  and therefore at most $\xi_\PBH/f_\NL^{2} \lesssim 10^{-2}$ within the interesting range of scales, generating any relevant clustering at formation, $\xi_\PBH\gtrsim 1$, would
 require $|f_\NL|\gsim 10$. Currently, the COBE/FIRAS limit 
$(\mu <9\cdot 10^{-5}$) \cite{FIRAS} {\rm constrain} $A_l\lsim 10^{-4}$, corresponding to a necessary value of $|f_\NL|\gsim 1$. It is  also interesting to notice that this estimate is consistent with the result reported in Ref.~\cite{Young:2019gfc}. Looking at their Fig.~6, we see that the merger rate is impacted by the NG corrections if $f_\NL \zeta_l \gtrsim 10^{-2}$, where $\zeta_l$ is the typical amplitude of the large-scale part of the curvature perturbation.  Using the maximum allowed value $\zeta_l \sim A_l^{1/2}\sim 10^{-2}$, one finds that clustering becomes more sizeable than the Poisson distribution precisely for $f_\NL \gtrsim 1$. 
Notice also that, as long as $\xi_{\PBH} \lesssim 1$, the overall PBH abundance is not altered by the NG since  the short-scale variance is significantly shifted  only for 
$f_\NL\gsim A_l^{-1/2}$.  This justifies  the use of the Gaussian formula to compute the abundance and, consequently, 
we have chosen the corresponding Gaussian value of the parameter $\nu$ to have $f_\PBH=10^{-3}$.
Notice that changing the abundance requires only a tiny change in the parameter $\nu$, since $f_\PBH $ is exponentially sensitive to $\nu$ as $f_\PBH \sim \exp(-\nu^2/2)$, and therefore to $A_s$~\cite{sasaki}, implying our conclusions are robust with respect to changes in the overall PBH abundance. 
{ Notice though that another source of non-Gaussianity is introduced by the unavoidable non-linear relation between the density contrast and the curvature perturbation~\cite{DeLuca:2019qsy, Young:2019yug}. This independent effect would modify the amplitude $A_s$ of a factor of order unity to maintain the same PBH abundance, without affecting our results. Furthermore, this ineludible NG is a small scale effect, and is not affected by the large-scale NG discussed in this paper.}

 Large PBH clustering will require large values of $|f_\NL|$. However, one may not consider such large values at will. As mentioned in the Introduction, the coupling between small and large scales introduces an isocurvature dark matter anisotropy from the PBHs in the CMB anisotropies which is severely constrained by Planck data. 
For the current lower bound $|f_\NL|\gsim 1$ from COBE/FIRAS to have large PBH clustering,  the isocurvature bound imposes $f_\PBH\lsim 5\cdot 10^{-4}$ \cite{tada,by,an}, making PBHs irrelevant as far as dark matter is concerned. Conversely, for large PBH abundances $f_\PBH=1$, the isocurvature bound imposes $|f_\NL|\lsim 4\cdot 10^{-4}$.
Of course, one can always envisage the situation in which the non-linear parameter $f_\NL$ is scale-dependent and switches on only at the scales relevant for the PBH binary formation and merger rates and dies off at the CMB scales, but we regard this possibility as rather artificial.

In the case in which the field $\chi(\vec x)$ introducing the large-scale PBH correlation is not the curvature perturbation, the forecasted limits on the CMB $\mu$-distortion in case of no detection will tell us that the PBHs may not be correlated at the time of formation.

Our results,  even though restricted to the standard and most studied formation mechanism of PBHs, interestingly  indicate that  future experiments looking for  CMB $\mu$-distortion  would constrain the hypothesis of PBH clustering at formation induced by local non-Gaussianity and would have a noticeable impact on the interpretation of the merger events seen so far  and on the possibility that PBHs in the LIGO/Virgo mass range may comprise the totality of the dark matter.
 The results discussed in this work may also extend the science case  supporting  future experiments aiming to constrain CMB $\mu$-distortions.
 Alternative scenarios for the formation of PBHs, such as through bubble collisions,  involve subhorizon dynamics, and, therefore, large-scale superhorizon clustering is not expected to arise.

\vskip 0.3cm
\noindent
\paragraph{Acknowledgments.}
\noindent
 V.DL., G.F. and 
A.R. are supported by the Swiss National Science Foundation 
(SNSF), project {\sl The Non-Gaussian Universe and Cosmological Symmetries}, project number: 200020-178787.

\bigskip



\begin{references}
	
	\bibitem{LIGOScientific:2018mvr}
	B.~P.~Abbott \textit{et al.} [LIGO Scientific and Virgo],
	Phys. Rev. X \textbf{9} (2019) no.3, 031040
	\arXiv{1811.12907}{astro-ph.HE}.
	

\bibitem{Abbott:2020niy}
R.~Abbott \textit{et al.} [LIGO Scientific and Virgo],
Phys. Rev. X \textbf{11} (2021), 021053
\arXiv{2010.14527}{gr-qc}.


\bibitem{Nakamura:1997sm}
T.~Nakamura, M.~Sasaki, T.~Tanaka and K.~S.~Thorne,
Astrophys. J. Lett. \textbf{487} (1997), L139-L142
\arXivold{astro-ph/9708060}.

\bibitem{Ioka:1998nz}
K.~Ioka, T.~Chiba, T.~Tanaka and T.~Nakamura,
Phys. Rev. D \textbf{58} (1998), 063003
\arXivold{astro-ph/9807018}.

\bibitem{Bird:2016dcv} 
S.~Bird, I.~Cholis, J.~B.~Munoz, Y.~Ali-Ha\"{i}moud, M.~Kamionkowski, E.~D.~Kovetz, A.~Raccanelli and A.~G.~Riess,
Phys.\ Rev.\ Lett.\  {\bf 116}, no. 20, 201301 (2016)
\arXiv{1603.00464}{astro-ph.CO}.


\bibitem{Sasaki:2016jop} 
M.~Sasaki, T.~Suyama, T.~Tanaka and S.~Yokoyama,
Phys.\ Rev.\ Lett.\  {\bf 117}, no. 6, 061101 (2016)
Erratum: [Phys.\ Rev.\ Lett.\  {\bf 121}, no. 5, 059901 (2018)]
\arXiv{1603.08338}{astro-ph.CO}.

\bibitem{Clesse:2016vqa}
S.~Clesse and J.~García-Bellido,
Phys. Dark Univ. \textbf{15} (2017), 142-147
\arXiv{1603.05234}{astro-ph.CO}.

\bibitem{Wang:2016ana}
S.~Wang, Y.~F.~Wang, Q.~G.~Huang and T.~G.~F.~Li,
Phys. Rev. Lett. \textbf{120} (2018) no.19, 191102
\arXiv{1610.08725}{astro-ph.CO}.

\bibitem{Ali-Haimoud:2017rtz} 
Y.~Ali-Ha\"{i}moud, E.~D.~Kovetz and M.~Kamionkowski,
Phys. Rev. D \textbf{96}, no.12, 123523 (2017)
\arXiv{1709.06576}{astro-ph.CO}.

\bibitem{Chen:2018czv}
Z.~C.~Chen and Q.~G.~Huang,
Astrophys. J. \textbf{864} (2018) no.1, 61
\arXiv{1801.10327}{astro-ph.CO}.

\bibitem{Kavanagh:2018ggo}
B.~J.~Kavanagh, D.~Gaggero and G.~Bertone,
Phys. Rev. D \textbf{98} (2018) no.2, 023536
\arXiv{1805.09034}{astro-ph.CO}.

\bibitem{raidal}
M.~Raidal, C.~Spethmann, V.~Vaskonen and H.~Veerm\"{a}e,
JCAP \textbf{02} (2019), 018
\arXiv{1812.01930}{astro-ph.CO}.


\bibitem{Fernandez:2019kyb}
N.~Fernandez and S.~Profumo,
JCAP \textbf{08} (2019), 022
\arXiv{1905.13019}{astro-ph.HE}.


\bibitem{ver}
V.~Vaskonen and H.~Veerm\"{a}e,
Phys. Rev. D \textbf{101} (2020) no.4, 043015
\arXiv{1908.09752}{astro-ph.CO}.



\bibitem{Gow:2019pok}
A.~D.~Gow, C.~T.~Byrnes, A.~Hall and J.~A.~Peacock,
JCAP \textbf{01} (2020), 031
\arXiv{1911.12685}{astro-ph.CO}.





\bibitem{Hall:2020daa}
A.~Hall, A.~D.~Gow and C.~T.~Byrnes,
Phys. Rev. D \textbf{102} (2020), 123524
\arXiv{2008.13704}{astro-ph.CO}.



\bibitem{DeLuca:2020jug}
V.~De Luca, V.~Desjacques, G.~Franciolini and A.~Riotto,
JCAP \textbf{11} (2020), 028
\arXiv{2009.04731}{astro-ph.CO}.

\bibitem{DeLuca:2020sae}
V.~De Luca, V.~Desjacques, G.~Franciolini, P.~Pani and A.~Riotto,
Phys. Rev. Lett. \textbf{126} (2021) no.5, 051101
\arXiv{2009.01728}{astro-ph.CO}.


\bibitem{Wong:2020yig}
K.~W.~K.~Wong, G.~Franciolini, V.~De Luca, V.~Baibhav, E.~Berti, P.~Pani and A.~Riotto,
Phys. Rev. D \textbf{103} (2021) no.2, 023026
\arXiv{2011.01865}{gr-qc}.


\bibitem{Hutsi:2020sol}
G.~H\"utsi, M.~Raidal, V.~Vaskonen and H.~Veerm\"ae,
JCAP \textbf{03} (2021), 068
\arXiv{2012.02786}{astro-ph.CO}.

\bibitem{us} V.~De Luca, G.~Franciolini, P.~Pani and A.~Riotto,
JCAP \textbf{05} (2021), 003
\arXiv{2102.03809}{astro-ph.CO}.
 
\bibitem{Franciolini:2021tla}
G.~Franciolini, V.~Baibhav, V.~De Luca, K.~K.~Y.~Ng, K.~W.~K.~Wong, E.~Berti, P.~Pani, A.~Riotto and S.~Vitale,
\arXiv{2105.03349}{gr-qc}.
 
 
\bibitem{Ali-Haimoud:2018dau}
Y.~Ali-Ha\"\i{}moud,
Phys. Rev. Lett. \textbf{121} (2018) no.8, 081304
\arXiv{1805.05912}{astro-ph.CO}.

\bibitem{in} V.~Desjacques and A.~Riotto,
Phys. Rev. D \textbf{98}, no.12, 123533 (2018)
\arXiv{180610414}{astro-ph.CO}.


\bibitem{Ballesteros:2018swv}
G.~Ballesteros, P.~D.~Serpico and M.~Taoso,
JCAP \textbf{10}, 043 (2018)
\arXiv{1807.02084}{astro-ph.CO}.

\bibitem{MoradinezhadDizgah:2019wjf}
A.~Moradinezhad Dizgah, G.~Franciolini and A.~Riotto,
JCAP \textbf{11} (2019), 001
\arXiv{1906.08978}{astro-ph.CO}.

\bibitem{Raidal:2017mfl}
M.~Raidal, V.~Vaskonen and H.~Veerm\"ae,
JCAP \textbf{09} (2017), 037
\arXiv{1707.01480}{astro-ph.CO}.

\bibitem{Young:2019gfc}
S.~Young and C.~T.~Byrnes,
JCAP \textbf{03} (2020), 004
\arXiv{1910.06077}{astro-ph.CO}.

\bibitem{atal} V.~Atal, A.~Sanglas and N.~Triantafyllou,
JCAP \textbf{11}, 036 (2020)
\arXiv{2007.07212}{astro-ph.CO}.

\bibitem{ng} N.~Bartolo, E.~Komatsu, S.~Matarrese and A.~Riotto,
Phys. Rept. \textbf{402}, 103-266 (2004)
\arXivold{astro-ph/0406398}.

\bibitem{sasaki} 
M.~Sasaki, T.~Suyama, T.~Tanaka and S.~Yokoyama,
Class. Quant. Grav. \textbf{35}, no.6, 063001 (2018)
\arXiv{1801.05235}{astro-ph.CO}.

\bibitem{Green:2020jor}
A.~M.~Green and B.~J.~Kavanagh,
J. Phys. G \textbf{48} (2021) no.4, 4
\arXiv{2007.10722}{astro-ph.CO}.

\bibitem{Jedamzik:2020ypm}
K.~Jedamzik,
JCAP \textbf{09} (2020), 022
\arXiv{2006.11172}{astro-ph.CO}.

\bibitem{Bringmann:2018mxj}
T.~Bringmann, P.~F.~Depta, V.~Domcke and K.~Schmidt-Hoberg,
Phys. Rev. D \textbf{99} (2019) no.6, 063532
\arXiv{1808.05910}{astro-ph.CO}.

\bibitem{Planck} 
Y.~Akrami \textit{et al.} [Planck],
Astron. Astrophys. \textbf{641}, A9 (2020)
\arXiv{1905.05697}{astro-ph.CO}.

\bibitem{tada}
Y.~Tada and S.~Yokoyama,
Phys. Rev. D \textbf{91}, no.12, 123534 (2015)
\arXiv{1502.01124}{astro-ph.CO}.



\bibitem{by} S.~Young and C.~T.~Byrnes,
JCAP \textbf{04}, 034 (2015)
\arXiv{1503.01505}{astro-ph.CO}.

\bibitem{an} N.~Bartolo, D.~Bertacca, V.~De Luca, G.~Franciolini, S.~Matarrese, M.~Peloso, A.~Ricciardone, A.~Riotto and G.~Tasinato,
JCAP \textbf{02}, 028 (2020)
\arXiv{1909.12619}{astro-ph.CO}.

\bibitem{Chluba:2012we}
J.~Chluba, A.~L.~Erickcek and I.~Ben-Dayan,
Astrophys. J. \textbf{758} (2012), 76
\arXiv{1203.2681}{astro-ph.CO}.

\bibitem{s} J.~Chluba and R.~A.~Sunyaev,
Mon. Not. Roy. Astron. Soc. \textbf{419}, 1294-1314 (2012)
\arXiv{1109.6552}{astro-ph.CO}.

\bibitem{Chluba:2015bqa}
J.~Chluba, J.~Hamann and S.~P.~Patil,
Int. J. Mod. Phys. D \textbf{24} (2015) no.10, 1530023
\arXiv{1505.01834}{astro-ph.CO}.

\bibitem{mu1} A.~Kogut, D.~J.~Fixsen, D.~T.~Chuss, J.~Dotson, E.~Dwek, M.~Halpern, G.~F.~Hinshaw, S.~M.~Meyer, S.~H.~Moseley and M.~D.~Seiffert, \textit{et al.}
JCAP \textbf{07}, 025 (2011)
\arXiv{1105.2044}{astro-ph.CO}.


\bibitem{mu2} J.~Chluba, M.~H.~Abitbol, N.~Aghanim, Y.~Ali-Haimoud, M.~Alvarez, K.~Basu, B.~Bolliet, C.~Burigana, P.~de Bernardis and J.~Delabrouille, \textit{et al.}
\arXiv{1909.01593}{astro-ph.CO}.

\bibitem{mu3} J. Chluba J. et al.,  BAAS \textbf{51} 184 (2019).

\bibitem{1967SvA....10..602Z}
Zel'dovich, Y.B. and Novikov, I.D.: 1967, {\it Soviet Astronomy} {\bf 10}, 602.

\bibitem{Hawking:1974rv}
S.~W.~Hawking,
Nature \textbf{248} (1974), 30-31


\bibitem{Chapline:1975ojl}
G.~F.~Chapline,
Nature \textbf{253}, no.5489, 251-252 (1975)

\bibitem{s1} P.~Ivanov, P.~Naselsky and I.~Novikov,
Phys.\ Rev.\ D {\bf 50}, 7173 (1994).

\bibitem{s00} 
S.~Blinnikov, A.~Dolgov, N.~K.~Porayko and K.~Postnov,
JCAP {\bf 1611}, 036 (2016)
\arXiv{1611.00541}{astro-ph.HE}.

\bibitem{Suyama:2019cst}
T.~Suyama and S.~Yokoyama,
PTEP \textbf{2019} (2019) no.10, 103E02
\arXiv{1906.04958}{astro-ph.CO}.

\bibitem{Musco:2018rwt}
I.~Musco,
Phys. Rev. D \textbf{100} (2019) no.12, 123524
\arXiv{1809.02127}{gr-qc}.

\bibitem{Germani:2018jgr}
C.~Germani and I.~Musco,
Phys. Rev. Lett. \textbf{122} (2019) no.14, 141302
\arXiv{1805.04087}{astro-ph.CO}.

\bibitem{Musco}
I.~Musco, V.~De Luca, G.~Franciolini and A.~Riotto,
Phys. Rev. D \textbf{103} (2021) no.6, 063538
\arXiv{2011.03014}{astro-ph.CO}.


\bibitem{Gow:2020bzo}
A.~D.~Gow, C.~T.~Byrnes, P.~S.~Cole and S.~Young,
JCAP \textbf{02} (2021), 002
\arXiv{2008.03289}{astro-ph.CO}.


\bibitem{Bardeen:1985tr}
J.~M.~Bardeen, J.~R.~Bond, N.~Kaiser and A.~S.~Szalay,
Astrophys. J. \textbf{304} (1986), 15-61

\bibitem{Inman:2019wvr}
D.~Inman and Y.~Ali-Ha\"\i{}moud,
Phys. Rev. D \textbf{100} (2019) no.8, 083528
\arXiv{1907.08129}{astro-ph.CO}.

 
\bibitem{Hu:1994bz}
W.~Hu, D.~Scott and J.~Silk,
Astrophys. J. Lett. \textbf{430} (1994), L5-L8
\arXivold{astro-ph/9402045}.

\bibitem{Dent:2012ne}
J.~B.~Dent, D.~A.~Easson and H.~Tashiro,
Phys. Rev. D \textbf{86}, 023514 (2012)
\arXiv{1202.6066}{astro-ph.CO}.


\bibitem{carr} 
B.~Carr, K.~Kohri, Y.~Sendouda and J.~Yokoyama,
\arXiv{2002.12778}{astro-ph.CO}.


\bibitem{FIRAS} D.~J.~Fixsen, E.~S.~Cheng, J.~M.~Gales, J.~C.~Mather, R.~A.~Shafer and E.~L.~Wright,
Astrophys. J. \textbf{473}, 576 (1996)
\arXivold{astro-ph/9605054} 


\bibitem{DeLuca:2019qsy}
V.~De Luca, G.~Franciolini, A.~Kehagias, M.~Peloso, A.~Riotto and C.~\"Unal,
JCAP \textbf{07} (2019), 048
\arXiv{1904.00970}{astro-ph.CO}.

\bibitem{Young:2019yug}
S.~Young, I.~Musco and C.~T.~Byrnes,
JCAP \textbf{11} (2019), 012
\arXiv{1904.00984}{astro-ph.CO}.



\end{references}
\end{document}